\documentclass[11pt,american]{article}
\usepackage[T1]{fontenc}
\usepackage[utf8]{inputenc}
\usepackage{babel}
\usepackage{array}
\usepackage{refstyle}
\usepackage{float}
\usepackage{booktabs}
\usepackage{units}
\usepackage{mathtools}
\usepackage{varwidth}
\usepackage{amsmath}
\usepackage{amssymb}
\usepackage{graphicx}
\usepackage[numbers]{natbib}
\usepackage{microtype}
\usepackage[pdfusetitle,
 bookmarks=true,bookmarksnumbered=false,bookmarksopen=false,
 breaklinks=false,pdfborder={0 0 1},backref=false,colorlinks=false]
 {hyperref}

\makeatletter

\AtBeginDocument{\providecommand\Figref[1]{\ref{Fig:#1}}}
\AtBeginDocument{\providecommand\Eqref[1]{\ref{Eq:#1}}}
\AtBeginDocument{\providecommand\Tabref[1]{\ref{Tab:#1}}}
\providecommand{\tabularnewline}{\\}

\RS@ifundefined{subsecref}
  {\newref{subsec}{name = \RSsectxt}}
  {}
\RS@ifundefined{thmref}
  {\def\RSthmtxt{theorem~}\newref{thm}{name = \RSthmtxt}}
  {}
\RS@ifundefined{lemref}
  {\def\RSlemtxt{lemma~}\newref{lem}{name = \RSlemtxt}}
  {}

\@ifundefined{date}{}{\date{}}
\usepackage{arxiv}

\usepackage{booktabs}
\usepackage{graphicx}
\usepackage{microtype}
\usepackage{tikz}
\usepackage{enumitem}
\usepackage{caption}
\usepackage{subcaption}
\usepackage{xcolor}
\usepackage{float}

\setlist{nosep}

\makeatother

\begin{document}
\title{Multi-Aperture PPG with MAPIS: Spatial Optical and Temporal Coherence
Fields}
\author{Shuguang Wang and Yuanjing Wang}
\maketitle
\begin{abstract}
Conventional reflectance photoplethysmography (PPG) typically reduces
tissue optical responses to one or a few detector channels. Matrix
Pinhole Image Sensing (MAPIS) instead simultaneously samples multiple
aperture-dependent optical signals. We analyzed 44 recordings of 30
seconds each from a single MAPIS device to characterize the spatial
relationships among DC intensity, pulsatile AC, perfusion index (PI),
and temporal coherence across Red and infrared channels. AC increased
sublinearly with DC at both wavelengths, resulting in higher PI toward
lower-DC apertures, consistent with a first-order relative path-sensitivity
interpretation. A second counterintuitive spatial trend was observed
in temporal coherence: lower-DC and lower-AC apertures exhibited higher
autocorrelation and greater cardiac-periodic spectral concentration.
IR signals also showed consistently higher temporal coherence than
corresponding Red signals. These observations demonstrate that multi-aperture
PPG resolves reproducible spatial relationships among optical intensity,
fractional pulsatile sensitivity, and waveform coherence that are
largely averaged together in conventional single-channel PPG. The
present study establishes within-device relationships and motivates
controlled studies of the underlying photon-path mechanisms.
\end{abstract}

\section{Introduction}

Photoplethysmography (PPG) measures optically induced changes associated
with the cardiac modulation of blood volume and is conventionally
decomposed into a pulsatile AC component superimposed on a slowly
varying DC baseline \citep{Allen2007}. In reflection geometry, however,
both components depend on wavelength, source--detector separation,
tissue optical properties, source emission pattern, detector acceptance,
and local tissue structure. Diffuse optical theory predicts that spatially
resolved reflectance contains information about absorption and transport
scattering \citep{Farrell1992,Durduran2010}, while Monte Carlo and
experimental PPG studies show that source--detector separation, wavelength,
and source/detector geometry alter photon path length, penetration
depth, DC level, and pulsatile response \citep{Chatterjee2019,SoleMorillo2025,ReiserGeometry2025}.

Spatially resolved diffuse reflectance has long been used to estimate
or constrain tissue optical properties, and average photon path length
can vary substantially with source--detector separation and optical
properties \citep{Nilsson2002}. These observations motivate a multi-path
interpretation of reflectance PPG: different apertures do not merely
provide replicated measurements with different gains, but may weight
different photon-path ensembles and therefore exhibit distinct fractional
sensitivities to physiological perturbations.

MAPIS (Matrix Pinhole Image Sensing) implements this idea using an
array of pinholes that simultaneously samples the returned optical
field. Each aperture yields both an aggregate temporal signal and
an image patch, potentially providing spatial, spectral, angular,
and temporal information. The present study focuses on the aggregate
temporal signals and asks three questions: (i) how DC, AC, and PI
vary across apertures; (ii) whether PI can be related to photon-path
sensitivity under a first-order optical model; and (iii) how cardiac-period
temporal coherence varies across apertures and wavelengths, and whether
its spatial organization is related to effective optical sampling.

\section{MAPIS Geometry and Data}

\subsection{Optical layout}

The MAPIS sensor contains a $4\times9$ aperture array with sub-millimeter-scale
aperture spacing and micrometer-scale aperture openings. Exact geometric
dimensions of the device used in this study are proprietary and are
therefore not disclosed. A Red LED (655 $nm$) and an infrared (IR)
LED (940 $nm$) are mounted beneath the aperture plane in a compact
shared package. Their emission centers are laterally separated by
approximately half a millimeter. The nominal vertical source-to-aperture
separation is approximately 3 mm, although the effective source position
and angular emission field may vary because of packaging and assembly
tolerances.

\begin{figure}[h]
\centering
\includegraphics[scale=0.75]{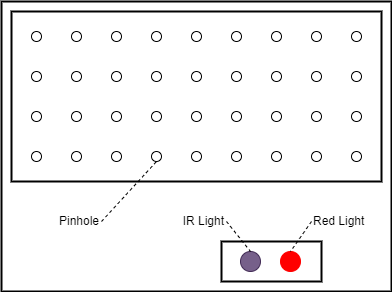}

\caption{Layout of the MAPIS aperture array and light sources. The Red (655
nm) and IR (940 nm) LEDs are located beneath the aperture array within
a compact shared package.}\label{fig:pinhole-lightsource-layout}
\end{figure}

\begin{figure}[h]
\centering
\includegraphics[scale=0.6]{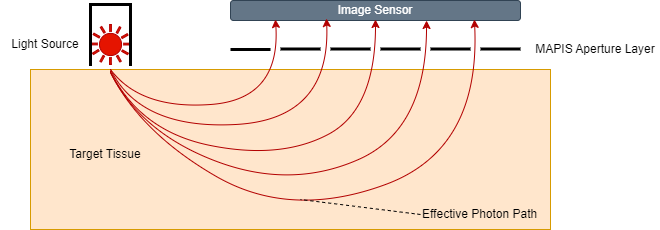}

\caption{Schematic cross-sectional view of the MAPIS optical geometry. Light
propagating through the target tissue is sampled by different imaging
apertures on the image sensor. The illustrated curves represent effective
photon-path ensembles rather than individual photon trajectories.}\label{fig:MAPIS-cross-sectional-view}
\end{figure}

\Figref{pinhole-lightsource-layout} and \Figref{MAPIS-cross-sectional-view}
illustrate the principal structure and optical geometry of the MAPIS
sensor.

\subsection{Dataset and analysis scope}

The analysis included 44 recordings, each 30 seconds in duration,
acquired on a single MAPIS device. Each recording contained signals
from 36 aperture channels at both Red and IR wavelengths, sampled
at 90 Hz. The present study therefore reports conditional observations
from a single device rather than population- or device-generalized
estimates.

For each aperture $i$ and wavelength $\lambda$, the following quantities
were derived: DC, peak-to-peak AC, perfusion index $PI=AC/DC$, one-cycle
normalized autocorrelation $ACF(T_{H})$, and the cardiac-periodic
power fraction $Q_{\mathrm{periodic}}$. RoR was defined from the
red-to-IR PI ratio.

\subsection{Signal processing and statistical analysis}

For each 30-s recording, Red and IR signals from all 36 apertures
were processed using the same pipeline. After discarding the first
90 frames, DC was calculated from the retained signal, AC from the
detrended pulsatile component, and $PI=AC/DC$. A record-common cardiac
period $T_{H}$ was estimated over a heart-rate search range of 30--180
beats/min and used for the one-cycle autocorrelation analysis.

For temporal-coherence analysis, signals were smoothed, detrended,
and temporally centered before calculating ACF at the cardiac lag.
$Q_{\mathrm{periodic}}$ was derived from Welch power spectra \citep{Proakis2021}
using harmonic bands centered at $f_{H}$, $2f_{H}$, and $3f_{H}$,
with half-width $\max(0.15\ \mathrm{Hz},\,0.15f_{H}),$and normalized
by total power from 0.1 to 45 Hz.

Spatial relationships across the 36 apertures were quantified using
Spearman rank correlation \citep{Weaver2017}, low-order radial and
two-dimensional spatial models, and mean absolute percentage error
(MAPE) on held-out data. The AC--DC exponent was estimated from
\begin{equation}
\log AC=\alpha+b\log DC.
\end{equation}

DC--PI spatial compensation was summarized by a gradient cancellation
fraction. Associations between ACF and $Q_{\mathrm{periodic}}$ were
evaluated after removal of record-level and fixed-aperture effects,
and matched Red--IR differences were summarized by their median.
Unless otherwise stated, summary statistics gave equal weight to each
recording.

\section{Optical Model}

\subsection{First-order observation model}

For aperture $i$ and wavelength $\lambda$, the measured intensity
is written as
\begin{equation}
I_{i,\lambda}(t)=G_{i,\lambda}R_{i,\lambda}(q(t)),
\end{equation}

where $G_{i,\lambda}$ is a time-invariant multiplicative instrument
factor over the analysis window, $R_{i,\lambda}$ is the tissue optical
response, and $q(t)=q_{0}+\delta q(t)$ is a physiological state variable.
A first-order expansion gives
\begin{equation}
I_{i,\lambda}(t)\approx G_{i,\lambda}R_{i,\lambda}(q_{0})+G_{i,\lambda}\left.\frac{\partial R_{i,\lambda}}{\partial q}\right|_{q_{0}}\delta q(t).
\end{equation}

Hence
\begin{align}
DC_{i,\lambda} & \approx G_{i,\lambda}R_{i,\lambda}(q_{0}),\\
AC_{i,\lambda} & \approx G_{i,\lambda}\left|\frac{\partial R_{i,\lambda}}{\partial q}\right|\left|\Delta q\right|,
\end{align}

and therefore
\begin{equation}
PI_{i,\lambda}\equiv\frac{AC_{i,\lambda}}{DC_{i,\lambda}}\approx\left|\frac{\partial\ln R_{i,\lambda}}{\partial q}\right|\left|\Delta q\right|.\label{eq:pi_jac}
\end{equation}

\Eqref{pi_jac} distinguishes mean optical level from fractional sensitivity:
DC and PI are not expected to share the same spatial dependence.

\subsection{Path-distribution interpretation of PI}

For an absorption perturbation, a path-ensemble representation may
be written as
\begin{equation}
R_{i,\lambda}(\mu_{a,\lambda})=\int_{0}^{\infty}W_{i,\lambda}(L)\exp(-\mu_{a,\lambda}L)\,dL,
\end{equation}

where $W_{i,\lambda}(L)$ is a non-negative weighting over detected
photon path lengths. If $W$ is approximately fixed during a small
absorption perturbation,
\begin{equation}
\frac{\partial\ln R_{i,\lambda}}{\partial\mu_{a,\lambda}}=-\frac{\int LW_{i,\lambda}(L)e^{-\mu_{a,\lambda}L}\,dL}{\int W_{i,\lambda}(L)e^{-\mu_{a,\lambda}L}\,dL}=-\langle L\rangle_{i,\lambda}^{\mathrm{det}}.
\end{equation}

The superscript “det” denotes a detection-weighted quantity, i.e.,
the average is taken over photon paths contributing to the detected
signal. Thus, under an absorption-dominant and spatially common pulsatile
perturbation,
\begin{equation}
PI_{i,\lambda}\approx\langle L\rangle_{i,\lambda}^{\mathrm{det}}\,\left|\Delta\mu_{a,\lambda}\right|.
\end{equation}

This does not make PI an absolute path length measurement; $W(L)$
itself depends on scattering, anisotropy, geometry, boundaries, source
emission, and detector acceptance \citep{Cheong1990,Jacques2013,Wang1995,Durduran2010}.

\subsection{Asymptotic AC--DC relation}

The full diffusion-dipole solution contains source- and image-source
terms of the form \citep{Farrell1992,Jacques2013}
\begin{equation}
\left(\mu_{\mathrm{eff}}+\frac{1}{r_{k}}\right)\frac{e^{-\mu_{\mathrm{eff}}r_{k}}}{r_{k}^{2}}.
\end{equation}

At sufficiently large source-detector separation, $r_{k}\approx r$,
yielding the asymptotic radial dependence
\begin{equation}
R(r)\sim C\frac{e^{-\mu_{\mathrm{eff}}r}}{r^{2}},\qquad\mu_{\mathrm{eff}}=\sqrt{3\mu_{a}(\mu_{a}+\mu_{s}')},
\end{equation}

with $\mu_{s}'=\mu_{s}(1-g)$. Therefore
\begin{equation}
\ln R=\ln C-\mu_{\mathrm{eff}}r-2\ln r.
\end{equation}

If the pulsatile perturbation acts primarily through a small change
in $\mu_{a}$, then
\begin{equation}
\frac{\partial\ln R}{\partial\mu_{a}}=-r\frac{\partial\mu_{\mathrm{eff}}}{\partial\mu_{a}}+\frac{\partial\ln C}{\partial\mu_{a}}.
\end{equation}

At sufficiently large $r$, provided that the absorption dependence
of $C$ does not introduce a term with comparable $r$-dependence,
the term proportional to $r$ dominates the derivative, giving
\begin{equation}
\left|\frac{\partial\ln R}{\partial\mu_{a}}\right|\propto r.
\end{equation}

Under the first-order relation
\begin{equation}
PI_{i,\lambda}\approx\left|\frac{\partial\ln R}{\partial\mu_{a}}\right|\left|\Delta\mu_{a}\right|,
\end{equation}

a spatially common pulsatile absorption perturbation therefore gives
\begin{equation}
PI\left(r\right)\propto r.
\end{equation}

Consequently, the spatial envelopes of DC and AC satisfy
\begin{equation}
DC\left(r\right)\propto\frac{e^{-\mu_{\mathrm{eff}}r}}{r^{2}},
\end{equation}
and
\begin{equation}
AC\left(r\right)=DC\left(r\right)PI\left(r\right)\propto\frac{e^{-\mu_{\mathrm{eff}}r}}{r}.\label{eq:AC_DC_PI_r}
\end{equation}

Thus, under these asymptotic assumptions, the AC envelope contains
one fewer inverse-distance factor than the DC envelope. Equivalently,
\begin{equation}
\frac{AC\left(r\right)}{DC\left(r\right)}\propto r,
\end{equation}
 so AC decreases more slowly with source--detector separation than
DC.

Let
\begin{equation}
b_{\mathrm{local}}=\frac{\nicefrac{d\ln AC}{dr}}{\nicefrac{d\ln DC}{dr}}=\frac{\mu_{\mathrm{eff}}r+1}{\mu_{\mathrm{eff}}r+2},
\end{equation}

and hence, for $\mu_{\mathrm{eff}}r>0$,
\begin{equation}
\frac{1}{2}<b_{\mathrm{local}}<1.
\end{equation}
We use this relation only as a qualitative mechanism for the observed
sublinear AC-DC scaling, not as an estimator of $\mu_{s}'$.

\section{Temporal Coherence Measures}

For a detrended waveform $x_{i,\lambda}(t)$, the normalized autocorrelation
is evaluated at the cardiac period $T_{H}=\nicefrac{1}{f_{H}}$:
\begin{equation}
ACF_{i,\lambda}(T_{H})=\frac{\langle x_{i,\lambda}(t),x_{i,\lambda}(t+T_{H})\rangle}{\|x_{i,\lambda}(t)\|\,\|x_{i,\lambda}(t+T_{H})\|}.\label{eq:acf_def}
\end{equation}

Pure amplitude scaling cancels in \Eqref{acf_def}, so ACF is not
an amplitude metric. We additionally define the cardiac-periodic power
fraction as
\begin{equation}
Q_{\mathrm{periodic}}=\frac{P_{f_{H}}+P_{2f_{H}}+P_{3f_{H}}}{P_{\mathrm{total}}},
\end{equation}

where $f_{H}=\nicefrac{1}{T_{H}}$, the numerator contains power in
predefined bands around the cardiac fundamental and its second and
third harmonics, and the denominator is the total signal power. By
Fourier-series representation, a periodic waveform with fundamental
frequency $f_{H}$ contains spectral components at integer harmonics
$nf_{H}$, while the Wiener--Khinchin theorem establishes that autocorrelation
and power spectral density are Fourier-transform pairs \citep{Oppenheim2016}.
Accordingly, $Q_{\mathrm{periodic}}$ is used here as a compact spectral
descriptor of cardiac-cycle coherence rather than an independent physiological
observable.

\section{Results}

\begin{figure}[h]
\centering
\includegraphics[scale=0.3]{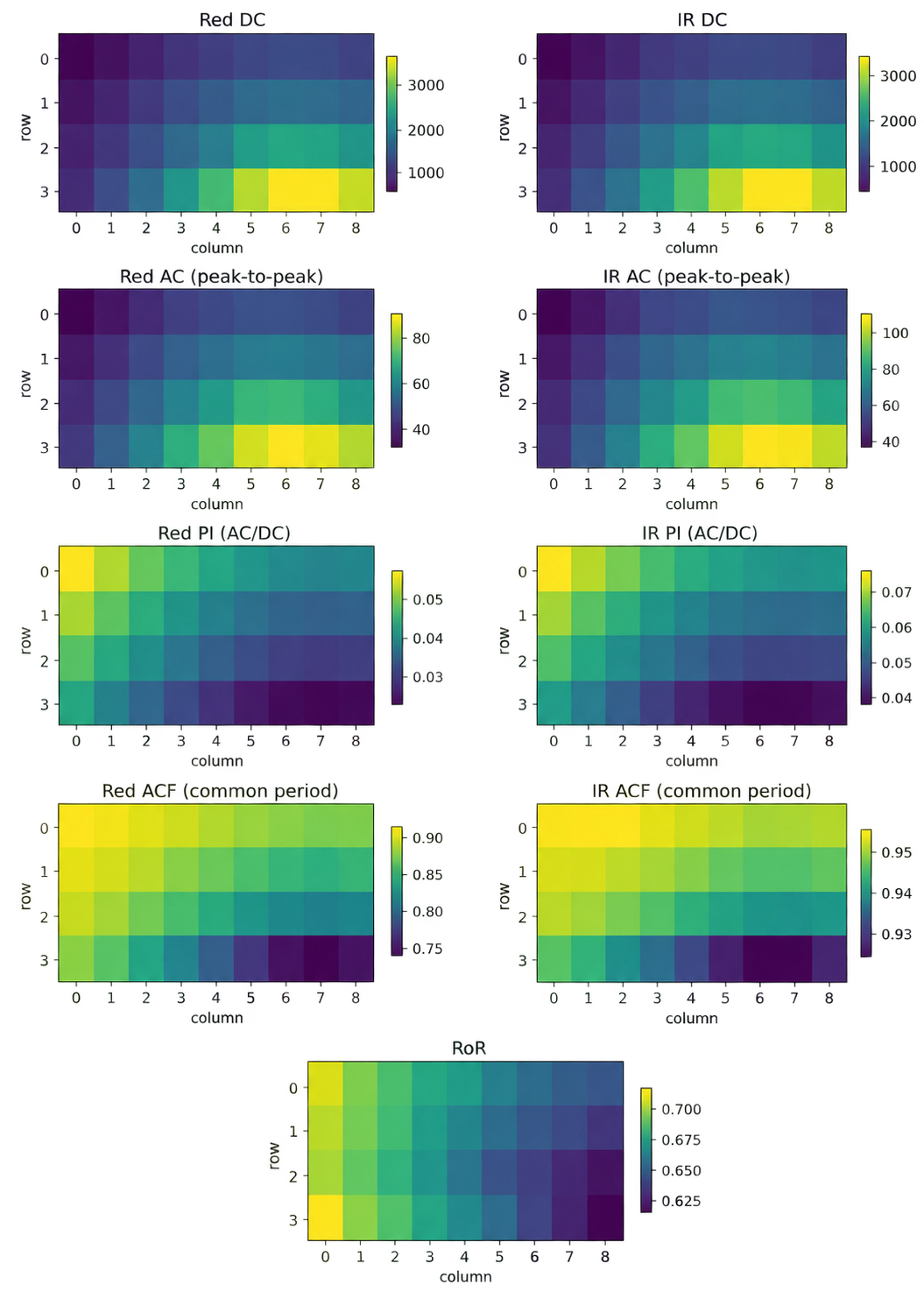}

\caption{Equal-record mean spatial maps obtained from 44 MAPIS recordings
of 30-seconds each. Despite the lower DC and AC levels in more weakly
illuminated apertures, PI and ACF increase in these regions, revealing
spatial relationships that are not evident from signal amplitude alone.
The distinct Red and IR patterns further indicate wavelength-dependent
optical transport. These maps illustrate spatial relationships that
are not directly accessible from a conventional single-channel PPG
measurement.}\label{fig:Equal-record-mean-spatial}
\end{figure}

The analysis included 44 MAPIS recordings, each 30 s in duration,
all acquired using a single device. \Figref{Equal-record-mean-spatial}
shows the equal-record mean spatial maps of DC, AC, PI, and ACF for
both the Red and IR channels, together with the corresponding RoR
map.

Several notable findings emerge:
\begin{itemize}
\item AC decreases more slowly with source--detector separation than DC.
Equivalently, PI tends to be higher in apertures with lower DC and
lower AC, consistent with the asymptotic optical model described above.
\item ACF also exhibits a stable spatial pattern, with higher values in
apertures with lower DC and lower AC. The optical origin of this relationship
remains unresolved.
\item For the same aperture, IR generally exhibits higher temporal coherence
than Red, while PI also shows a strong wavelength-dependent spatial
structure.
\end{itemize}
The following sections describe these findings in greater detail.

\subsection{DC, AC, and PI form distinct spatial fields}

Across the 44 recordings, brighter apertures generally exhibited larger
absolute AC but smaller PI. Empirically,
\begin{equation}
AC\propto DC^{b},
\end{equation}
with record-level median exponents $b=0.6135$ for Red and $b=0.6220$
for IR. All 44 records exhibited $b<1$ at both wavelengths. Thus
AC grows sublinearly with DC, and PI decreases toward brighter channels.

The spatial PI field was strongly structured across the aperture array
and inversely related to DC. In record-wise Spearman analysis, the
median $\rho(DC,PI)$ was approximately $-0.977$ for Red and $-0.959$
for IR. Radial distance from the nominal source location captured
much of the Red ordering, with median $\rho(r,PI)\approx0.962$, but
was less adequate for IR, where $\rho(r,PI)\approx0.692$. This lower
radial correlation should not be interpreted as weak spatial organization
of the IR PI field; rather, it indicates that the IR field contains
a substantial non-radial component.

This interpretation is supported by held-out spatial modeling. A radial-only
PI model gave mean absolute percentage errors of approximately $6.48\%$
for Red and $14.79\%$ for IR. A low-order two-dimensional quadratic
spatial model reduced these errors to approximately $5.15\%$ and
$6.23\%$, respectively, and a model combining log-DC with the two-dimensional
spatial terms further reduced them to approximately $4.34\%$ for
Red and $5.04\%$ for IR.

PI should therefore be interpreted as a two-dimensional relative path-sensitivity
field rather than as a simple function of radial source-detector distance,
or as a direct measurement of geometric or absolute photon path length.
In particular, the effective optical field may deviate from an ideal
point-source geometry because of LED package structure, angular emission,
assembly tolerances, and other fixed source-detector response effects.
The stronger non-radial component observed in IR is consistent with
this broader optical-field interpretation.

\subsection{PI partially compensates the DC spatial gradient}

Using a common low-order spatial representation for DC and PI, the
PI gradient was consistently directed opposite to the DC gradient.
The median cancellation fraction was approximately $36.7\%$ for Red
and $35.6\%$ for IR. Thus, part of the spatial decrease in DC is
compensated by an increase in PI, quantitatively explaining why AC
decreases more slowly across the aperture field than DC. This behavior
is consistent with the sublinear AC--DC relation described above
and does not require the weaker AC signals at lower-DC apertures to
be attributed to noise.

\subsection{ACF exhibits a counterintuitive spatial trend}

A counterintuitive but reproducible observation was that apertures
with lower DC and lower AC often exhibited higher ACF and higher $Q_{\mathrm{periodic}}$.
This spatial trend is qualitatively similar to that of PI, even though
ACF and PI represent different signal properties. After removal of
record-level and fixed-aperture effects, ACF remained strongly correlated
with $Q_{\mathrm{periodic}}$, with correlations of approximately
$0.984$ for Red and $0.982$ for IR, confirming that the observed
ACF field reflects changes in cardiac-periodic waveform coherence
rather than a simple amplitude effect.

Simple fixed additive-noise models predict the opposite tendency:
lower-amplitude signals should generally exhibit poorer periodic coherence.
The observed increase in ACF and $Q_{\mathrm{periodic}}$ toward lower-DC
and lower-AC apertures therefore suggests that the spatial pattern
is associated with differences in the underlying optical sampling
rather than ordinary amplitude-limited SNR.

One possible interpretation is that apertures closer to the light
source sample a more localized photon-path ensemble and are therefore
more sensitive to local, nonperiodic tissue perturbations. Apertures
sampling longer or more spatially distributed effective photon paths
may average over a larger tissue volume, reducing the contribution
of local perturbations relative to the cardiac-periodic component
and thereby increasing ACF and $Q_{\mathrm{periodic}}$. This interpretation
remains a working hypothesis and requires independent experimental
validation.

\subsection{Stable wavelength dependence of temporal coherence}

For every matched aperture-record observation in the present dataset
$(44\times36=1584\:pairs)$, IR ACF exceeded Red ACF. The median difference
was
\begin{equation}
ACF_{IR}-ACF_{R}\approx0.087.
\end{equation}

IR also exhibited higher $Q_{\mathrm{periodic}}$, lower spectral
entropy, and more repeatable beat morphology. Spectral reconstruction
indicated that most of the Red--IR coherence difference arose from
off-harmonic low-to-mid-frequency content rather than from simple
high-frequency white noise. The optical or physiological origin of
this wavelength dependence remains unresolved.

One possible hypothesis is that the higher IR coherence arises from
a mechanism similar to the spatial ACF trend described above. Because
IR light generally penetrates more deeply and may sample a more spatially
distributed photon-path ensemble, its effective path distribution
may provide stronger averaging of local, non-cardiac tissue perturbations.
This could reduce off-harmonic variability and increase the relative
coherence of the cardiac-periodic component. This interpretation remains
a working hypothesis and requires independent experimental validation.

\subsection{Summary of principal observations}

\Tabref{Principal-Observations} summarizes the principal findings.

\begin{table}[H]
\centering
\caption{Principal Observations}\label{tab:Principal-Observations}

\begin{tabular}{V{\linewidth}V{\linewidth}}
\toprule 
\addlinespace
\textbf{Observation} & \textbf{Interpretation}\tabularnewline\addlinespace
\midrule
\addlinespace
{\footnotesize AC decreases more slowly than DC }{\footnotesize\par}

{\footnotesize across the aperture field} & {\footnotesize PI increases toward lower-DC apertures, consistent with}{\footnotesize\par}

{\footnotesize the asymptotic optical model.}\tabularnewline\addlinespace
\midrule
\addlinespace
{\footnotesize PI forms a structured two-}{\footnotesize\par}

{\footnotesize dimensional spatial field} & {\footnotesize PI reflects relative path sensitivity rather than simple
radial}{\footnotesize\par}

{\footnotesize distance or absolute photon path length.}\tabularnewline\addlinespace
\midrule
\addlinespace
{\footnotesize PI partially opposes the DC spatial}{\footnotesize\par}

{\footnotesize gradient} & {\footnotesize This compensation explains the sublinear AC--DC }{\footnotesize\par}

{\footnotesize relationship.}\tabularnewline\addlinespace
\midrule
\addlinespace
{\footnotesize Lower-DC and lower-AC apertures}{\footnotesize\par}

{\footnotesize exhibit higher temporal coherence} & {\footnotesize Temporal coherence appears to depend on optical sampling, }{\footnotesize\par}

{\footnotesize not simply on signal amplitude or additive-noise SNR.}\tabularnewline\addlinespace
\midrule
\addlinespace
{\footnotesize IR exhibits consistently higher }{\footnotesize\par}

{\footnotesize temporal coherence than Red} & {\footnotesize A possible explanation is stronger spatial averaging
of local}{\footnotesize\par}

{\footnotesize non-cardiac perturbations by the IR photon-path ensemble.}\tabularnewline\addlinespace
\bottomrule
\end{tabular}

\end{table}

\section{Discussion}

\subsection{Relation to spatially resolved diffuse optics and multi-distance
PPG}

The present findings are consistent with the broader diffuse-optics
literature, in which spatially resolved reflectance and source--detector
separation encode information about optical transport \citep{Farrell1992,Durduran2010,Nilsson2002}.
They are also consistent with PPG studies showing that source--detector
separation and sensor geometry materially affect DC level, optical
path length, penetration depth, and the relative pulsatile component
\citep{Chatterjee2019,SoleMorillo2025,ReiserGeometry2025}. 

A distinctive aspect of MAPIS is the simultaneous dense sampling of
many aperture-dependent channels under the same physiological state.
This enables direct comparison of steady optical intensity, fractional
pulsatile sensitivity, and temporal coherence across different effective
photon-path ensembles without sequential sensor repositioning.

\subsection{A multi-layer interpretation of MAPIS observables}

The data support a hierarchical interpretation in which DC represents
the mean optical transport field, PI represents a conditional fractional-sensitivity
field, AC is the product of these two quantities, and ACF together
with $Q_{\mathrm{periodic}}$ characterizes the temporal coherence
of the dynamic signal component. These observables may share spatial
structure, but they represent different aspects of the underlying
optical and physiological process and should not be treated as interchangeable.

Pure multiplicative gain cancels from both PI and normalized ACF.
By contrast, source geometry, angular emission, detector acceptance,
additive offsets, detector nonlinearity, and path re-weighting can
modify the effective observation kernel and thereby alter the measured
spatial fields.

\subsection{Path-ensemble interpretation of temporal coherence}

The spatial and wavelength-dependent ACF trends may reflect a common
path-ensemble mechanism. Both lower-DC/farther apertures and IR illumination
are expected to sample photon populations with different spatial and
depth weighting than brighter/nearer apertures or Red illumination.
If these path ensembles average local non-cardiac tissue perturbations
more effectively, the cardiac-periodic component would occupy a larger
fraction of the observed dynamic signal, leading to higher ACF and
$Q_{\mathrm{periodic}}$.

This interpretation provides a unified explanation for the aperture-dependent
and wavelength-dependent coherence fields, but it remains non-unique.
A stronger test would require controlled perturbations that independently
modify optical path distributions, tissue motion, contact conditions,
or scattering while preserving the underlying cardiac rhythm.

\subsection{Implications for scattering-sensitive sensing}

The path-ensemble framework suggests that changes in tissue absorption
or scattering may reshape the spatial distributions of DC, PI, and
intra-aperture intensity. Because MAPIS simultaneously samples multiple
effective photon-path families, it may provide a useful platform for
studying transport-sensitive perturbations. Quantitative attribution
to absorption or scattering, however, requires independent optical
perturbations or phantoms with known $\mu_{a}$ and $\mu_{s}'$. The
present measurements do not uniquely identify either coefficient and
should therefore not be interpreted as direct measurements of tissue
scattering.

\section{Limitations}

This study is exploratory and focuses specifically on the spatial
relationships among AC, DC, PI, ACF, and related observables across
the MAPIS aperture array. For this reason, the present analysis uses
44 recordings acquired from a single device, so that cross-device
variations in source geometry, detector response, package optics,
and calibration do not confound the spatial relationships of interest.
The present results should therefore be interpreted as reproducible
within-device relationships rather than population-level physiological
laws or cross-device calibration results.

The source geometry is not perfectly represented by an ideal point-source
model, and the IR field contains substantial non-radial structure.
The path-distribution interpretation of PI further assumes a small,
approximately absorption-dominant perturbation with approximately
fixed path weighting. The relationship between ACF and $Q_{\mathrm{periodic}}$
is partly mathematical because both are derived from the same waveform.
Finally, the present data do not uniquely identify absolute $\mu_{a}$,
$\mu_{s}'$, photon path length, tissue sampling depth, $SpO\ensuremath{_{2}}$,
or glucose.

\section{Conclusion}

MAPIS reveals reproducible aperture-dependent spatial relationships
among DC, AC, PI, and temporal-coherence measures that cannot be described
as fixed gain-scaled copies of a single PPG waveform. DC, PI, and
AC exhibit distinct but coupled spatial fields, with PI partially
compensating the DC gradient and thereby producing the observed sublinear
AC--DC relationship. ACF and $Q_{\mathrm{periodic}}$ show a separate
but similarly structured spatial trend, with higher temporal coherence
in lower-DC and lower-AC apertures. In addition, IR channels consistently
exhibit higher temporal coherence than corresponding Red channels.

These findings suggest that effective photon-path ensembles and wavelength
materially influence the spatial and temporal structure of pulsatile
optical signals. The present single-device study establishes reproducible
within-device relationships rather than cross-device or population-level
laws. Multi-device validation, characterization of source and detector
response, and controlled optical-phantom experiments will be required
to determine which MAPIS observables provide robust and independently
identifiable transport- or scattering-sensitive information.

\section*{Ethics Statement}

The measurements analyzed in this study were obtained from adult company
personnel who voluntarily participated in non-invasive fingertip optical
measurements conducted as part of internal device development and
engineering evaluation. The analyzed data were de-identified. The
original data collection was not conducted under a formal institutional
research protocol.

\section*{Patent Notice}

MAPIS is a patented technology owned by TechInu. Certain aspects of
the MAPIS technology described in this article may be protected by
patents.

\bibliographystyle{plainnat_arXiv}
\bibliography{mapis_formal_paper_ppg_optics}

@article{Allen2007,
  author = {Allen, John},
  title = {Photoplethysmography and its application in clinical physiological measurement},
  journal = {Physiological Measurement},
  year = {2007}, volume = {28}, number = {3}, pages = {R1--R39},
  doi = {10.1088/0967-3334/28/3/R01}, pmid = {17322588}
}

@article{Cheong1990,
  author = {Cheong, W. F. and Prahl, S. A. and Welch, A. J.},
  title = {A review of the optical properties of biological tissues},
  journal = {IEEE Journal of Quantum Electronics},
  year = {1990}, volume = {26}, number = {12}, pages = {2166--2185},
  doi = {10.1109/3.64354}
}

@article{Jacques2013,
  author = {Jacques, Steven L.},
  title = {Optical properties of biological tissues: a review},
  journal = {Physics in Medicine and Biology},
  year = {2013}, volume = {58}, number = {11}, pages = {R37--R61},
  doi = {10.1088/0031-9155/58/11/R37}, pmid = {23666068}
}

@article{Farrell1992,
  author = {Farrell, Thomas J. and Patterson, Michael S. and Wilson, Brian},
  title = {A diffusion theory model of spatially resolved, steady-state diffuse reflectance for the noninvasive determination of tissue optical properties in vivo},
  journal = {Medical Physics},
  year = {1992}, volume = {19}, number = {4}, pages = {879--888},
  doi = {10.1118/1.596777}, pmid = {1518476}
}

@article{Durduran2010,
  author = {Durduran, Turgut and Choe, Regine and Baker, Wesley B. and Yodh, Arjun G.},
  title = {Diffuse optics for tissue monitoring and tomography},
  journal = {Reports on Progress in Physics},
  year = {2010}, volume = {73}, number = {7}, pages = {076701},
  doi = {10.1088/0034-4885/73/7/076701}
}

@article{Wang1995,
  author = {Wang, Lihong and Jacques, Steven L. and Zheng, Liqiong},
  title = {MCML---Monte Carlo modeling of light transport in multi-layered tissues},
  journal = {Computer Methods and Programs in Biomedicine},
  year = {1995}, volume = {47}, number = {2}, pages = {131--146},
  doi = {10.1016/0169-2607(95)01640-F}, pmid = {7587160}
}

@article{Nilsson2002,
  author = {Nilsson, Henrik and Larsson, Marcus and Nilsson, Gert E. and Str{\"o}mberg, Tomas},
  title = {Photon pathlength determination based on spatially resolved diffuse reflectance},
  journal = {Journal of Biomedical Optics},
  year = {2002}, volume = {7}, number = {3}, pages = {478--485},
  doi = {10.1117/1.1482378}, pmid = {12175300}
}

@article{Chatterjee2019,
  author = {Chatterjee, Subhasri and Kyriacou, Panayiotis A.},
  title = {Monte Carlo Analysis of Optical Interactions in Reflectance and Transmittance Finger Photoplethysmography},
  journal = {Sensors},
  year = {2019}, volume = {19}, number = {4}, pages = {789},
  doi = {10.3390/s19040789}, pmid = {30769957}, pmcid = {PMC6412556}
}

@article{SoleMorillo2025,
  author = {Sol{\'e} Morillo, {\'A}ngel and Lambert Cause, Joan and De Pauw, Kevin and da Silva, Bruno and Stiens, Johan},
  title = {Exploring Near- and Far-Field Effects in Photoplethysmography Signals Across Different Source--Detector Distances},
  journal = {Sensors},
  year = {2025}, volume = {25}, number = {1}, pages = {99},
  doi = {10.3390/s25010099}, pmid = {39796889}, pmcid = {PMC11722670}
}

@article{ReiserGeometry2025,
  author = {Reiser, Maximilian and Mueller, Timm and Breidenassel, Andreas and Amft, Oliver},
  title = {Source-Detector Geometry Analysis of Reflective PPG by Measurements and Simulations},
  journal = {IEEE Open Journal of Engineering in Medicine and Biology},
  year = {2025}, volume = {6}, pages = {400--406},
  doi = {10.1109/OJEMB.2025.3546771}, pmid = {40657054}, pmcid = {PMC12251171}
}

@book{Oppenheim2016,
  author    = {Oppenheim, Alan V. and Verghese, George C.},
  title     = {Signals, Systems and Inference},
  publisher = {Pearson},
  year      = {2016},
  isbn      = {9780133943283}
}

@book{Proakis2021,
  author    = {Proakis, John G. and Manolakis, Dimitris G.},
  title     = {Digital Signal Processing: Principles, Algorithms and Applications},
  edition   = {5},
  publisher = {Pearson},
  year      = {2021},
  isbn      = {9780137348657}
}

@book{Weaver2017,
  author    = {Weaver, Kathleen F. and Morales, Vanessa and Dunn, Sarah L.
               and Godde, Kanya and Weaver, Pablo F.},
  title     = {An Introduction to Statistical Analysis in Research:
               With Applications in the Biological and Life Sciences},
  publisher = {Wiley},
  year      = {2017},
  doi       = {10.1002/9781119454205},
  isbn      = {9781119299684}
}

\end{document}